\documentclass{emulateapj}
\usepackage{epsf}
\usepackage{graphicx}

\newcommand{\etal}{{et al.~}}
\newcommand{\masyr}{ \ {\rm{mas \ yr^{-1}}}\>}
\newcommand{\kms}{ \ {\rm{km \ s^{-1}}}\>}

\begin{document}

\pagenumbering{arabic}
\title{Revisiting the Role of M31 in the Dynamical History of the
Magellanic Clouds}
\author{Nitya Kallivayalil\altaffilmark{1,2}}
\affil{MIT Kavli Inst. for Astrophysics \& Space Research, 70 Vassar
  Street, Cambridge, MA 02139}
\altaffiltext{1}{Pappalardo Fellow}
\altaffiltext{2}{nitya@mit.edu}
\author{Gurtina Besla}
\affil{Harvard-Smithsonian Center for Astrophysics, 60 Garden Street, 
Cambridge, MA 02138}
\author{Robyn Sanderson}
\affil{MIT Kavli Inst. for Astrophysics \& Space Research, 70 Vassar
  Street, Cambridge, MA 02139}
\author{Charles Alcock}
\affil{Harvard-Smithsonian Center for Astrophysics, 60 Garden Street, 
Cambridge, MA 02138}
%\altaffiltext{3}{calcock@cfa.harvard.edu}

\begin{abstract}
We study the dynamics of the Magellanic Clouds in a model for the
Local Group whose mass is constrained using the timing
argument/two-body limit of the action principle. The goal is to
evaluate the role of M31 in generating the high angular momentum orbit
of the Clouds, a puzzle that has only been exacerbated by the latest
$HST$ proper motion measurements. We study the effects of varying the
total Local Group mass, the relative mass of the Milky Way and M31,
the proper motion of M31, and the proper motion of the LMC on this
problem. Over a large part of this parameter-space we find that tides from
M31 are insignificant. For a range of LMC proper motions approximately
$3\sigma$ higher than the mean and total Local Group mass $> 3.5\times
10^{12} M_\odot$, M31 can provide a significant torque to the LMC
orbit. However, if the LMC is bound to the MW, then M31 is found to
have negligible effect on its motion and the origin of the high
angular momentum of the system remains a puzzle. Finally, we use the
timing argument to calculate the total mass of the MW-LMC system based
on the assumption that they are encountering each other for the first
time, their previous perigalacticon being a Hubble time ago, obtaining
$M_{\rm MW} + M_{\rm LMC} = (8.7 \pm 0.8) \times 10^{11} M_\odot$.
\end{abstract}


\section{Introduction}

The three-dimensional velocities of the Magellanic Clouds, from proper
motion measurements using $HST's$ Advanced Camera for Surveys (ACS)
and a sample of background QSOs (Kallivayalil \etal 2006a,b, hereafter
K1 and K2; see also Piatek \etal 2008, Kallivayalil \etal 2009), are
$\sim 100 \kms$ higher than those used in past theoretical modeling of
the Magellanic Stream (MS) and, for the LMC, now approach the escape
velocity of the Milky Way (MW). Consequently, as shown by Besla \etal
(2007), in a $\Lambda$CDM-based model for the MW the Clouds, assuming
they form a bound system, are likely on their first passage. This
claim has re-ignited the discussion about the origin of the
Clouds. Did they form in the outer regions of the Local Group or as
satellite galaxies of the MW?  Proposals for the latter, i.e., that
bind the Clouds to the MW, include the explicit use of smaller
velocities, for example, by giving the LMC \& SMC a common halo (Bekki
2008) as well as Modified Newtonian Dynamics (MOND) gravity (Wu \etal
2008). A recent study by Shattow \& Loeb (2009) argues that the past
orbit of the LMC is naturally confined within the virial radius of the
MW if a $\sim 14$\% increase in the MW circular velocity (Reid \&
Brunthaler 2004; Reid \etal 2009) is taken in combination
with the lower end of the allowed proper motion error-space.

Apart from uncertainties in the MW potential, other puzzles remain,
such as the high angular momentum of the LMC system (e.g. Fich \&
Tremaine 1991; Gott \& Thuan 1978). Even with the modest values for
transverse velocity ($\sim 100 \kms$) and apogalacticon distance (200
kpc) used in past MS models, a mass of $L_{\rm LMC} = 2 \times 10^{10}
M_\odot$ gives the following for the orbital angular momentum, $L_{\rm
LMC}$, of the LMC: $L_{\rm LMC} \sim M_{\rm LMC} R_{\rm LMC} V_ {\rm
LMC} \sim 4 \times 10^{14} M_\odot {\rm kpc} \kms$. This is equivalent
to the spin angular momentum of the Galactic disk, $L_{\rm disk} \sim
(2/3) M_{\rm disk} R_{\rm disk} V_0 \sim 4 \times 10^{14} M_\odot {\rm
kpc} \kms$, with standard values of $ M_{\rm disk} = 2 \times 10^{11}
M_\odot$, $R_{\rm disk} = 15$ kpc and $V_0 = 220 \kms$ (Sawa \&
Fujimoto 2005). Since the LMC is in a roughly polar orbit, these
angular momenta make right angles to each other.

This problem has seen many past iterations: Besla \etal (2007) show
that the LMC may be on a parabolic orbit, and hence the `angular
momentum problem' is moot. However, in this case an explanation of why
the LMC is moving so quickly and on such a different orbit from the
other satellites is warranted. For example, Fich \& Tremaine (1991)
comment that the maximum line-of-sight velocity of the MS ($-410
\kms$; Br{\"u}ns \etal 2005) is not only higher than that of the
Clouds themselves ($262 \kms$) but also higher than any MW satellite
within 200 kpc. Morphological studies of the satellite populations of
the MW and M31 arrive at a similar impasse. Van den Bergh (2006)
argues that the fact that the Clouds are gas-rich dIrr galaxies and
yet occupy the small Galactocentric distances usually dominated by
dSph galaxies may be accounted for by assuming that they are
interlopers that were originally formed in the outer reaches of the
Local Group.

Raychaudhury \& Lynden-Bell (1989) contended that M31 was close enough
to the LMC in its early orbital history to cause a significant tide,
and that this tide was oriented such as to generate the high orbital
angular momentum of the Clouds. It is worth noting here that their
analysis was done in the context of external tides on the MW-M31
system from more distant galaxies. Building on this theory, Shuter
(1992) and Byrd \etal (1994) considered the possibility that the
Clouds underwent a close encounter with M31 having only recently been
captured by the MW.  Given our new velocities it is especially
pertinent to not consider the LMC-SMC-MW system in isolation and to
explore whether the Clouds may have been subject to external torques
before entering the environs of the MW.  We revisit this classic
problem of quantifying the torque provided by M31 using the framework
of the timing argument/two-body limit of the least action principle,
given two new pieces of information: the new proper motions for the
Clouds \textit{and} a new transverse velocity estimate for M31 from
satellite velocities (van der Marel \& Guhathakurta 2008; hereafter
VG08).

Shattow \& Loeb (2009) also included the effect of M31 in their
analysis of whether the LMC is bound to the MW. We confirm the results
in \S~4 of their paper but take a different approach here. Within the
framework of the timing argument/two-body limit of the least-action
principle, could M31 have generated the high transverse velocities of
the Clouds?  There are 4 main effects on the Clouds' orbits that bear
exploring: 1) the effect of total Local Group mass, $M_{tot}$; 2) the
effect of $f=M_{\rm M31}/M_{\rm MW}$; 3) the effect of M31 proper
motion; and 4) the effect of varying the proper motion of the
LMC. There is a large range in values for $M_{tot}$ because this is a
quantity that cannot be directly measured without modeling. Various
methods give a range of $2 \times 10^{12} - 5.6 \times 10^{12}
M_\odot$ (Kochanek 1996; Wilkinson \& Evans 1999; Klypin \etal 2002;
VG08). The quantity $f$ is also hard to constrain observationally and
we look at a range of values from 0.8 - 2.0 (Einasto \& Lynden-Bell
1982; Evans \& Wilkinson 2000; Klypin \etal 2002; VG08). Recent VLBA
measurements might favor $f$ closer to unity (Reid \etal 2009). The
proper motion of M31 has not been directly measured, but as mentioned
above, indirect estimates exist and indicate that it is small
(VG08). We investigate the full range of values from the
literature. Finally, the proper motion of the LMC is the only
observationally well-constrained quantity in this analysis (K1) and we
test the whole error-space using a Monte-Carlo distribution.

The primary goal is to assess whether the inclusion of M31 in the
equations of motion of the Clouds can generate a significant torque on
the L/SMC orbit. In addition, we keep track of the magnitude of the
relative tidal force on the LMC from the MW and M31 to quantify
whether there are orbits in which the M31 tide is higher.  The tide is
calculated simply via a double radial differentiation of the potential
at the location of the L/SMC. While the locations of the
center-of-mass of the L/SMC are allowed to move under the influence of
the two more massive bodies, the potential shapes in all simulations
are kept fixed. Motivated by the findings of this analysis, we apply
the timing argument to the MW and LMC, assuming that the LMC is on its
first passage with the previous perigalacticon being a Hubble time,
$t_H$, ago, to calculate the mass of the MW.

%In B07 we discussed the finding that the two components of the proper
%motion control different orbital parameters of the LMC; namely
%$\mu_W$, the west-component of the proper motion, sets the tangential
%velocity and thus the orbital period and apogalacticon distance, while
%$\mu_N$, the north-component of the proper motion, governs the
%alignment of the past orbit of the LMC with the location of the MS in
%projection. Our value for $\mu_N$ predicts a deviation of the LMC's
%past orbit (by $\sim 7^\circ$) from the current location of the MS in
%the plane of the sky. All current MS models have assumed
%co-location. We attempted to use this deviation to distinguish between
%MW models and study MW halo-asphericity, and concluded that the
%deviation was independent of the model. Here we go one step further
%and investigate the effects of the `Local Group' on this
%deviation.

In most of the subsequent analysis we present results only for the LMC
and intend our conclusions to be representative of the LMC-SMC system
as a whole. It remains a possibility that the Clouds are not in a
binary system (K2; Piatek \etal 2008).  We do not explicitly
investigate this further as it is still possible to find a binary
orbit in the error space of the SMC for every given LMC orbit. Also, a
chance three-body interaction (MW-LMC-SMC) seems highly unlikely. It
has been speculated, for instance, that the Clouds are members of a
small subgroup that was captured by the Local Group (Metz \etal 2009),
or that fell into the MW at late times (D'Onghia \& Lake 2009). Given
their mass ratio ($\sim 10:1$) the SMC is not expected to play a major
role in shaping the orbit of the LMC. We thus carry out the analysis
assuming that the orbital path of the Clouds will be dependent on that
of the LMC, but we do include the perturbative effect of the SMC on
the LMC given the SMC mean velocity in all our calculations (note,
however, that this is an orbit in which the Clouds are unbound to
each other).  We await smaller proper motion errors for the SMC (see
\S~\ref{conclusion}) to lift these assumptions.

%%XXX
In \S~\ref{LGmodel} we recount briefly the work on the relative motion
of the MW and M31 and use this as the motivation for our Local Group
model and methods. In \S~3 we present results including changes to the
orbital history of the LMC given these new models. We also present the
results for MW mass. We conclude in \S~\ref{conclusion}.

\section{A Model for the Local Group}\label{LGmodel}
\subsection{The Relative Motion of the Milky Way and M31}\label{relmotion}
The Local Group is thought to be decoupled from the cosmological
expansion and gravitationally bound. This is supported by the fact
that its two major constituents, the MW and M31, are seen to be
approaching each other with a radial velocity of $\sim117 \kms$
(Binney \& Tremaine 1987). The tangential motion of M31 has not been
directly measured, but indirect estimates (VG08; Loeb \etal 2005;
Peebles \etal 2001) indicate that this tangential motion is small, and
thus support the assumption that the Local Group is bound.

Some simple dynamical arguments about the the Local Group can be made
based on known facts. In 1959, Kahn \& Woltjer proposed the
`timing argument', in which the MW and M31 first moved apart due to
general cosmological expansion, subsequently reversed paths and are
now falling into each other under their own gravitational attraction,
their motions being governed by Newtonian dynamics. Assuming a zero
angular momentum orbit (zero tangential motion) and the known current
separation and radial velocity, the timing argument requires a mass in
the Local Group $> 3 \times 10^{12} M_\odot$ to fulfill this
trajectory. An alternative explanation would be that the MW and M31
are accidentally passing by, but if the galaxies have moved at a
constant speed for a Hubble time, their separation would have changed
only by $\sim 117 \kms /H_0 \sim 1.6$ Mpc, i.e., less than the distance to the
next largest galaxy. So this is not a promising alternative (Peebles
1993).

A more sophisticated treatment of the dynamics of the Local Group,
pioneered by Peebles (1989; 1990; Shaya \etal 1995; Peebles \etal
2001), involves using the principle of least-action to calculate
orbital solutions from incomplete phase-space information by assuming
homogeneity of the early universe (see also Goldberg \& Spergel 2000;
Goldberg 2001). Note, however, that the two-body limit to this
solution is equivalent to the timing argument, since the equation of
relative motion has the same form as the cosmological acceleration
equation in the case of the evolution of a homogeneous, isotropic mass
distribution (Peebles 1993).

The timing argument has been investigated and applied widely in the
literature (Mishra 1985; Raychaudhury \& Lynden-Bell 1989; Kroeker \&
Carlberg 1991; Goldberg 2001). Two recent extensions of the formalism
bear further comment: (1) Chernin \etal (2009) include the antigravity
effect of dark energy in the Kahn-Woltjer model and find a larger
Local Group mass than in traditional methods of $4.5 \times 10^{12}
M_\sun$; (2) VG08 applied the timing argument to obtain a total mass
for the Local Group, after first deducing a transverse velocity for
M31 from satellite velocities. The methods they use for the latter are
a statistical analysis of the line-of-sight velocities of 17 M31
satellites, a study of the proper motions of two satellites M33 and IC
10 (from Brunthaler \etal 2005, 2007), and an analysis of the
line-of-sight velocities of 5 galaxies near the Local Group
turn-around radius. A full Monte Carlo analysis of all the
uncertainties involved produces a value of $V_{\rm tan} = 41.7 \kms$
for the median of the probability distribution of the Galactocentric
tangential velocity of M31, with $1 \sigma$ confidence interval
$V_{\rm tan} \leq 56 \kms$ (thus the radial orbit applied by Kahn \&
Woltjer is allowed in their solution). The inferred $1 \sigma$
confidence interval around the median Local Group mass obtained from
application of the timing argument is $5.58^{+0.85}_{-0.72} \times
10^{12} M_\sun$.

While this mass is consistent with most applications of the timing
argument, it is on the high end of what is predicted from $\Lambda$CDM
motivated galaxy models (Klypin \etal 2002; Li \& White 2008). Using
theoretical constraints to narrow the large space of solutions allowed
by uncertainties in satellite velocities and in halo extent, the
Klypin \etal models for M31 and the MW have a favored total mass of
$2.6 \times 10^{12} M_\sun$ (with a maximum upper-limit of roughly
twice this amount). The reason for the discrepancy is unclear, the
general wisdom being that the timing argument provides too simplistic
a view of mass-accretion. However, some studies (Kroeker \& Carlberg
1991) have investigated its accuracy in a cosmological context and
found it an adequate approximation to the true mass.

Our aim is not to build a `true' Local Group but rather to
investigate maximal M31 models, i.e., models in which M31 can hope to
generate large torques on the LMC orbit. Thus the timing argument
provides a natural framework in which to carry out these investigations.

\subsection{Methods}
\begin{figure}
\begin{center}
\includegraphics[scale=0.7]{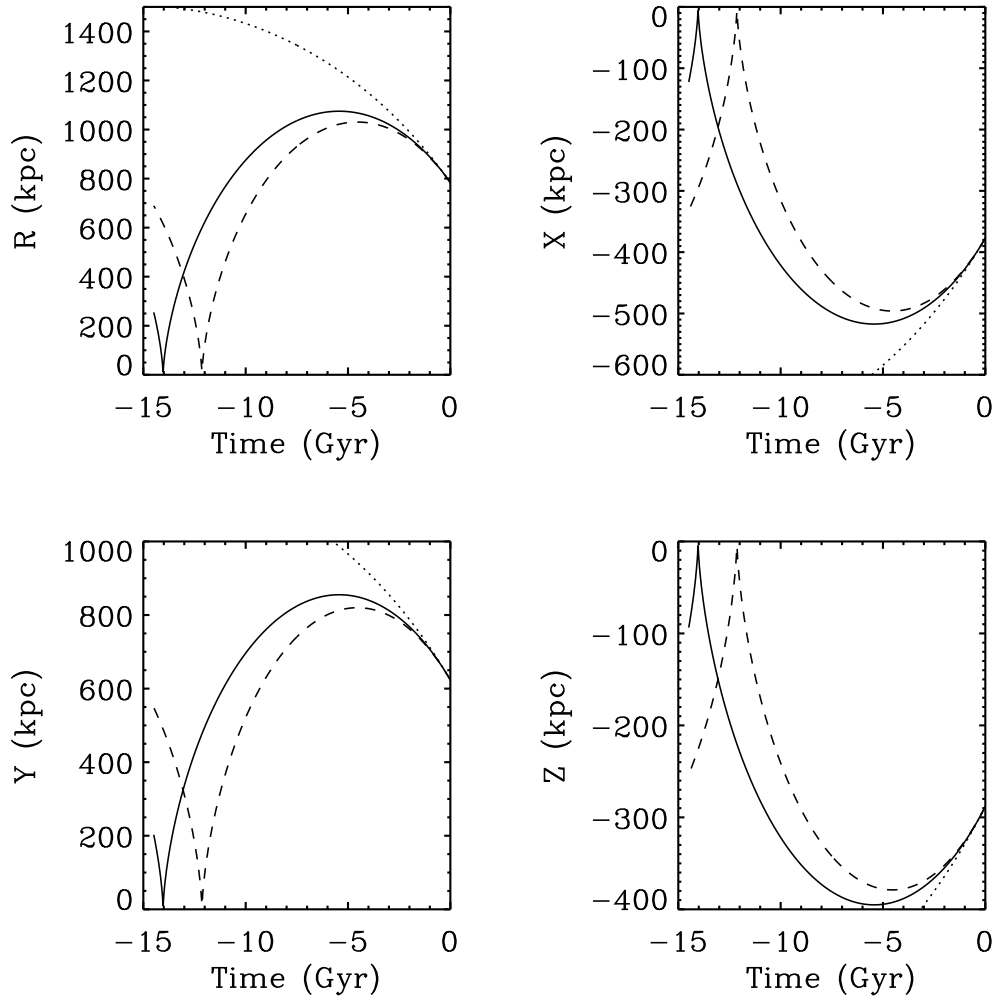}
\caption{\singlespace The relative orbit of MW and M31 is shown in our
Galactocentric $X,Y,Z$ reference frame for a few cases of total mass
$\mu/G$. The dotted line shows the past orbit for $\mu/G = 2.6 \times
10^{12} M_\odot$, the dashed line for $\mu/G = 5.2 \times 10^{12}
M_\odot$ and the solid line for $\mu/G = 4.6 \times 10^{12}
M_\odot$. }
\label{rM31MW}
\end{center}
\end{figure}

We estimate the effect of M31 by building on the timing argument
and considering some representative cases of the entire paramater-space:
we vary the total Local Group mass, $M_{tot}$, from $2.6-6 \times
10^{12} M_\sun$ and also consider two cases for the proper motion of
M31 - the mean value from VG08 as well as a radial orbit. Table~1
summarizes the quantities varied, the step-size of the variation or
error-space explored. We solve the two-body problem for the motion of
M31 relative to the MW in the same MW-centered coordinate system
($X,Y,Z$) that we used for the Clouds (see K1 and van der Marel \etal
2002) and then include M31's potential in the equations of motion for
the Clouds, knowing the distance between the Clouds and M31 at every
time-step. In solving for the radial orbit we assume a distance
modulus of 24.47 (McConnachie \etal 2005; Ribas \etal 2005), a radial
velocity (of M31 relative to the MW) of $-117 \kms$ (Binney \&
Tremaine 1987) and a current location for M31 of (RA = $10.68^\circ$,
Dec = $41.27^\circ$, J2000.0).  For the non-radial case we simply
adopt the six phase-space parameters supplied by VG08. Given the
adopted initial conditions, we are free to choose the total mass
(represented here as $\mu = G (M_{\rm M31} + M_{\rm MW}$)) so as to
specify the time in the past at which the MW and M31 were
co-located. Figure~\ref{rM31MW} shows the resulting motion of M31 with
respect to the MW in our Galactocentric frame for a few different
values of $\mu/G$ and a radial orbit.

As expected, orbits with total mass $\mu/G < 3 \times 10^{12} M_\odot$
do not `turn around' within $t_H$. In Figure~\ref{rM31MW} we show that
as $\mu$ increases, the turn around time decreases; the dotted line
shows the past orbit for $\mu/G = 2.6 \times 10^{12} M_\odot$, the
solid line shows the case of $\mu/G = 4.6 \times 10^{12} M_\odot$ and
the dashed line for $\mu/G = 5.2 \times 10^{12} M_\odot$. The
non-radial orbits have the same qualitative behavior with mass and we
do not plot them here. As in VG08, all quantities are found to vary
monotonically with the tangential velocity of M31: larger values of
tangential velocity lead to larger values of total mass, the period of
the orbit and the pericentric distance. We show the full Hubble time
evolution of the orbit in Figure~\ref{rM31MW} simply as a heuristic
exercise. There is little value in extending our static analysis to
significantly earlier times than 5 Gyr ago since stellar ages imply
that the Galactic disk (and presumably the halo) had not been fully
assembled at $z \ga 2$ (Wyse 2007; Cox \& Loeb 2008).

Once the mutual separation of M31 and MW is known it is easy to
introduce a potential term for M31 in the equations of motion for the
L/SMC:
\begin{equation}
\frac{d^2\mathbf{r}_{L}}{dt^2} =
\begin{array}{l}
\frac{\partial}{\partial\mathbf{r}_{L}}
[\phi_S(\mid\mathbf{r}_{L} - \mathbf{r}_{S}\mid) + \\
\phi_{\rm MW}(\mid\mathbf{r}_{L}\mid) + \phi_{\rm M31}(\mid\mathbf{r}_{L} - 
\mathbf{r}_{M31}\mid)] + \frac{{\rm F}_{L}}{M_L},
\end{array}
\end{equation}
where $\mathbf r_L$ is the Galactocentric distance of the LMC,
$\mathbf r_S$ that of the SMC and $\mathbf r_{M31}$ that of
M31. $\phi_S$ is the potential of the SMC, $\phi_{\rm MW}$ is the
potential of the MW and $\phi_{\rm M31}$ is that of M31. $F_L$ is the
dynamical friction on the LMC orbit and $M_L$ is the mass of the
LMC. There is an equivalent equation for the SMC.  In this model for
the Local Group, the MW and M31 are both approximated as isothermal
spheres with $\phi_{\rm MW, M31}(r) = -V_0^2 \ \ln r$, $V_{0, {\rm
MW}} = 220 \kms$ and $V_{0, {\rm M31}} = 250 \kms$ (Loeb \etal
2005). To be consistent with the total mass used in our 2-body
formulation we introduce a cut-off radius, $r_h$, outside which the
density drops to zero (see van der Marel \etal 2002):
\begin{equation}
\rho  \ = 
\left\{
\begin{array}{cc}
V_0^2/4 \pi Gr^2, & {\rm if} \ r \leq r_h, \\
0, & {\rm if} \ r > r_h.
\end{array}
\right.
\end{equation}
The enclosed mass is:
\begin{equation}
M(r)  \ = 
\left\{
\begin{array}{cc}
rV_0^2/G, & {\rm if} \ r \leq r_h, \\
r_hV_0^2/G \equiv M_{\rm tot}, & {\rm if} \ r > r_h. \\
\end{array}
\right.
\end{equation}

%\begin{eqnarray}
%\rho  \ = & V_0^2/4 \pi Gr^2, & {\rm if} \ r \leq r_h, \nonumber\\
%        = & 0, & {\rm if} \ r > r_h. \nonumber\\
%\end{eqnarray}
%The enclosed mass is:
%\begin{eqnarray}
%M(r)  \ = & rV_0^2/G, & {\rm if} \ r \leq r_h, \nonumber\\
%        = & r_hV_0^2/G \equiv M_{\rm tot}, & {\rm if} \ r > r_h. \nonumber\\
%\end{eqnarray}

The LMC and SMC are represented using Plummer models:
\begin{equation}
\phi_{L,S}(r) = GM_{L,S}/[(\mathbf{r} - \mathbf{r}_{L,S})^2 + K^2_{L,S}]^{1/2},
\end{equation}
with effective radii ($K_L, K_S$) of 3 and 2 kpc, and masses ($M_L,
M_S$) of $2 \times 10^{10} M_\odot$ and $3 \times 10^{9} M_\odot$,
respectively.  The final term, $\frac{{\rm F}_{L}}{M_L}$, accounts for
dynamical friction on the LMC orbit resulting from its motion through
the MW dark halo, for which we use the Chandrasekhar formula (Binney
\& Tremaine 1987):
\begin{equation}
\begin{array}{l}
F_{L,S} = \\
-\frac{4\pi G^2 M^2_{L,S} \ln(\Lambda) \rho(r_{L,S})}{v_{L,S}^3}
\left\{[{\rm erf}(X) - \frac{2X}{\sqrt{\pi}}e^{-X^2}]
\mathbf{v}_{L,S}\right\},
\end{array}
\end{equation}
where $X = v_{L,S}/V_{0, \rm MW}$ and $\rho(r_{L,S})$ is the density
of the MW halo at the Galactocentric distance of the L/SMC. The
strength of this force depends on the mass of the L/SMC, and since we
are integrating backwards in time from the current positions and
velocities, the sign of the dynamical friction term is that of
acceleration (see Figure~2). Our results do not depend sensitively on
the form of this drag because as shown in \S~3.1 it does not alter the
L/SMC orbit in such a way as to bring them significantly closer to M31
in the past. Our conclusions don't change even if it is
ignored. Further, we do not include dynamical friction from M31 in our
calculations because, as discussed in the next section, the LMC's
closest approach to M31 in the past 5 Gyr is roughly 500 kpc. Thus
dynamical friction from M31 is negligible.

%It is worth clarifying that the shapes of the MW, M31, LMC \& SMC do
%not deform in this analysis. Rather, the problem is the integration of
%pseudo point-particles in fixed potentials. The tidal force is
%represented by the gradient of the gravitational force at the position
%of the LMC in the limit that it has constant size. The goal is to see
%if the LMC comes close enough to M31 to make tidal torques important.

\section{Results}\label{orbitalhistory}
\subsection{LMC Orbits}
We use a Monte Carlo scheme to randomly draw 20,000 LMC proper motion
components from the errors in K1. From these initial values we allow
the LMC orbit to propagate backward in time in the Local Group model
described in \S~\ref{LGmodel}. The SMC proper motion is kept fixed at
its mean value (K2). We keep track of the orbits that take the LMC
closest to and furthest from M31.

As laid out in the introduction, we explore the effects of 1) $M_{
tot}$; 2) $f$; 3) M31 proper motion; and 4) LMC proper motion on the
past orbital history of the Clouds.  It was found that as $M_{tot}$
increases, the orbital period and apogalacticon of the LMC orbit is
smaller, as one might expect. However, $M_{tot}$ does not have any direct
bearing on the relative importance of M31 in the LMC's past. The
question becomes, given a value of $M_{tot}$, what are the salient
changes to the LMC orbit from varying $f$, the LMC proper motion and
M31's proper motion? The largest change to the orbit comes from
varying the LMC proper motion, followed by changes in $f$. The proper
motion of M31, by contrast, does not dramatically alter the
picture. It acts much like $M_{tot}$ : larger values of M31 proper
motion require larger values of $M_{tot}$ to fulfill the trajectory of
the timing argument. The currently accepted error-space of M31's
proper motion does not alter its relative distance to the LMC
dramatically. However, the proper motion of the LMC does.

\begin{figure}[!ht]
\begin{center}
\includegraphics[scale=0.7]{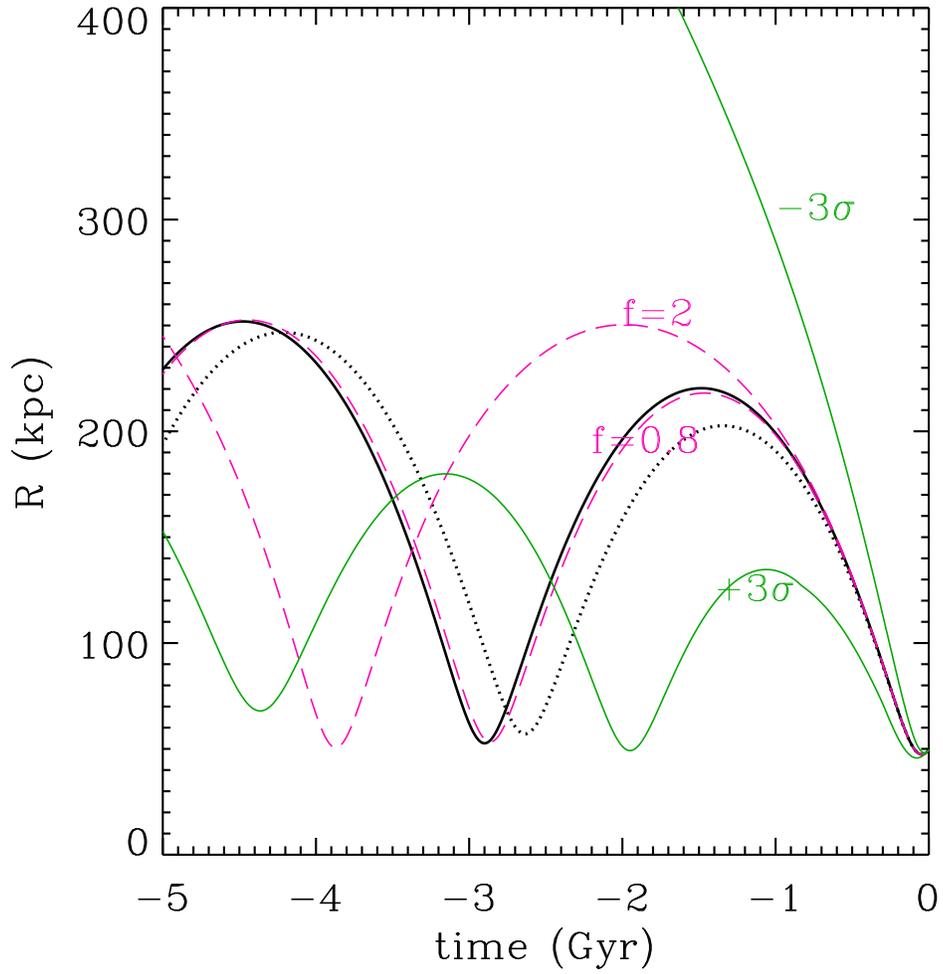}
\caption{\singlespace The Galactocentric distance of the LMC when M31
is included in the Local Group model ($f=1$; black solid line) and when it is
not (black dotted line). For a given $M_{tot}$, the pink lines mark
the effect of a maximum M31 model ($f=2$) and a minimum M31 model
($f=0.8$). The green lines mark the LMC orbit that comes closest to
M31, $\sim 3 \sigma$ increase in $\mu_W$ (marked $-3 \sigma$), and the
furthest orbit, $3 \sigma$ decrease in $\mu_W$ (marked $+3 \sigma$).}
\label{2bodyresults}
\end{center}
\end{figure}

We summarize the results in Figure~\ref{2bodyresults}.  The black
solid line shows the orbit with $M_{tot} = 5.6 \times 10^{12}
M_\odot$, $f=1$, mean LMC proper motion and mean M31 proper motion
(from VG08). This can be compared to the black dotted line which is
the orbit obtained in a MW-only Local Group (with $M_{\rm MW} = 2.8
\times 10^{12} M_\odot$). Note that this MW differs from our `infinite
mass' fiducial model in K2 in that it is modeled with a cut-off radius
and thus fixed total mass. We realize this is a larger MW mass than
indicated by either observations or $\Lambda$CDM modeling but
emphasize that we are trying to understand the relative importance of
M31 in LMC dynamics for which $M_{tot}$ itself turns out to be
unimportant. Varying $M_{tot}$ pushes the LMC orbit to larger or
shorter period, but the qualitative picture remains. As for the choice
of mass distribution, we contend that the treatment of M31 as an
isothermal sphere is adequate because, as described below, the LMC
does not come closer than $\sim 500$ kpc. As for the choice of mass
distribution for the MW, we showed in Figure~8 in Besla \etal (2007)
that the trajectories of the LMC line up in projection regardless of
whether an isothermal sphere or NFW (Navarro \etal 1996; 1997) profile
are used. Thus, replacing the isothermal spheres here with NFW models
is equivalent to looking at lower values of $M_{tot}$: the LMC would
effectively escape sooner from the MW in the past and with lower
velocity, but at the same time, M31 would be further away.

\begin{figure}[!ht]
\begin{center}
\includegraphics[scale=0.5]{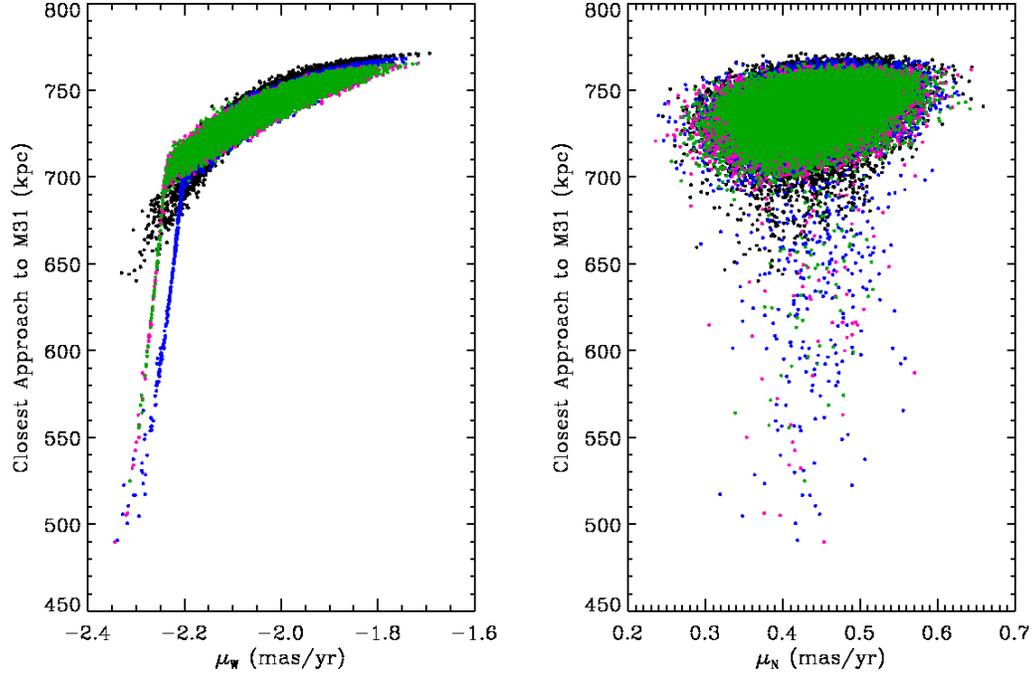}
\caption{\singlespace The closest approach to M31 as a function of the
  LMC's $\mu_W$ (\textit{left}) and $\mu_N$ (\textit{right}). The
  black dots show the case of $M_{tot} = 2.6\times10^{12} M_\odot$,
  the green dots show $M_{tot} = 3.5\times10^{12} M_\odot$, the blue
  dots show $M_{tot} = 4.6\times10^{12} M_\odot$ and the pink dots
  show $M_{tot} = 5.6\times10^{12} M_\odot$.  }
\label{fitparams}
\end{center}
\end{figure}

The dashed pink lines in Figure~\ref{2bodyresults} mark the boundaries
of $f=0.8$ (minimal M31 model) and $f=2.0$ (maximal M31 model) for
fixed $M_{tot}$. The green lines enclose the space of solutions given
by LMC proper motion: the LMC orbits with closest approach to M31 have
the mean western LMC proper motion, $\mu_W$, in the range of
$\mu_W-(>3\sigma)$, while the furthest orbits are for
$\mu_W+(>3\sigma)$. These also naturally correspond to unbound and
bound orbits to the MW, respectively. The northern component of the
proper motion, $\mu_N$, does not affect the trajectory as dramatically
(see Figure~\ref{fitparams}).

Within a period of 5 Gyr, the closest orbits bring the LMC within
$\sim 500$ kpc of M31, while the furthest orbits come within $\sim
770$ kpc.  Since the distance to M31 is relatively large throughout
the duration of our calculations and tides drop off as $r^{-3}$ it
seems unlikely that M31 would have a pronounced effect on the orbits
of the Clouds, and this is the case for most of the parameter-space that
we search.  However, for a range of LMC proper motions $\mu_W -
(>3\sigma)$ the torque provided by M31 can be significant. In
Figure~\ref{fitparams} we show the closest distance to M31 reached by
the LMC as a function of the LMC's $\mu_W, \mu_N$, and their
corresponding error circles. The black dots are for $M_{tot} =
2.6\times10^{12} M_\odot$, the green dots for $M_{tot} =
3.5\times10^{12} M_\odot$, the blue dots for $M_{tot} =
4.6\times10^{12} M_\odot$ and the pink dots for $M_{tot} =
5.6\times10^{12} M_\odot$. The M31 tangential velocity is that of
VG08. There is a clear dependence on $\mu_W$ but not on $\mu_N$. For
the case of $-2.3 < \mu_W < -2.2 \masyr$ and $M_{tot} >
3.5\times10^{12} M_\odot$, the final relative velocity between the LMC
and M31 is smaller than the escape velocity of M31, but the relative
distance is larger than $r_{h, \rm M31}$. As $M_{tot}$ increases the
value of $\mu_W$ must also in general increase for this to be the
case. Varying $f$ broadens the distributions in Figure~\ref{fitparams}
(\textit{left} panel) by decreasing (high $f$) or increasing (low $f$)
the distances of closest approach.

\begin{figure}[!ht]
\begin{center}
\includegraphics[scale=0.5]{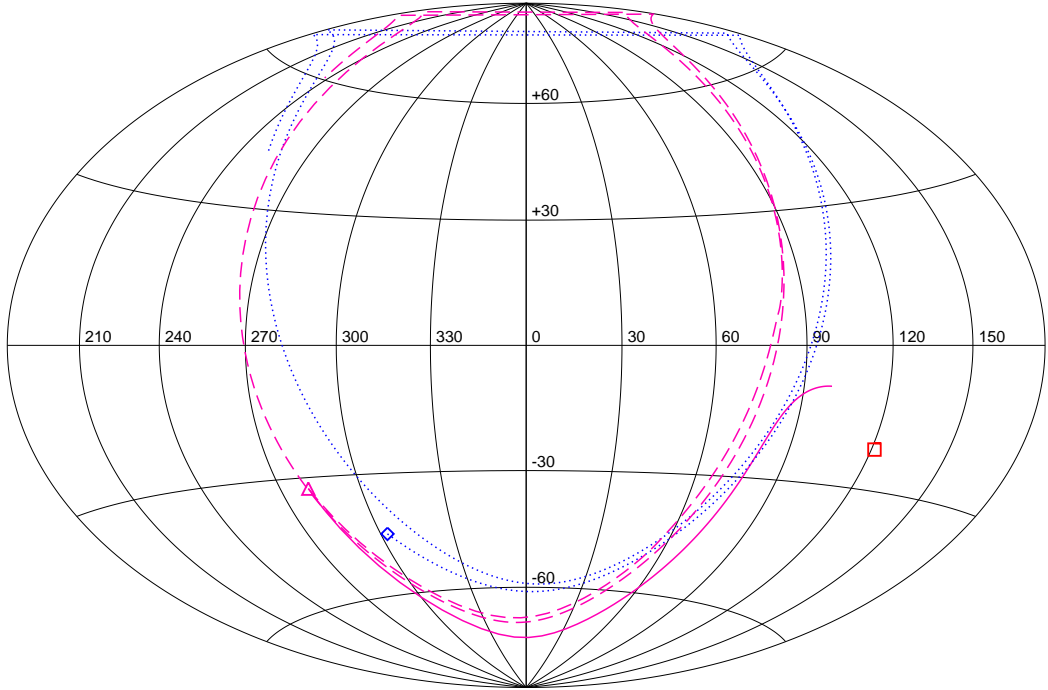}
\caption{\singlespace The orbital trajectory of the LMC and SMC in
  aitoff projection for the past 5 Gyr, centered on the MW. The pink
  triangle, blue diamond and red square mark the current positions of
  the LMC, SMC and M31 respectively, i.e., at $t=0$. The pink solid
  line shows the LMC orbit trajectory that takes it closest to M31 in
  the past, while the pink dashed lines show the corresponding
  furthest orbits. The blue dotted line shows the mean SMC orbit.}
\label{aitoff}
\end{center}
\end{figure}

Figure~\ref{aitoff} shows the past orbits of the Clouds in aitoff
projection, centered on the MW. The pink triangle, blue diamond and
red square mark the current positions of the LMC, SMC and M31
respectively. Again, the M31 tangential velocity is that of VG08. We
do not show the M31 orbit here because, as in the case of the radial
orbit, the trajectory in aitoff projection is roughly constant. For a
given $M_{tot}$ the pink solid line shows the LMC orbit trajectory
that takes it closest to M31, while the pink dashed lines show the
corresponding furthest orbits. The blue dotted line shows the mean SMC
orbit for reference. As seen in Figure~\ref{2bodyresults}, M31 does
provide a small amount of tidal pull on the LMC orbit compared to a
MW-only Local Group even for the mean velocities. However, M31 can
only \textit{alter} the orbit of the LMC if the LMC is moving roughly
$3-4 \sigma$ faster than we think, i.e., it is not bound to the MW at
all. For these cases the tidal influence of M31 is larger than that of
the MW (solid pink line) and the final relative velocity between the
LMC and M31 is lower than the escape velocity of M31, even though the
relative distance is larger than $r_{h, \rm M31}$. If we do allow our
calculations of these particular orbits to run for $t_H$ (not shown
here), the LMC eventually becomes bound to M31. However, accounting
for the hierarchical build-up of the Local Group, this analysis
becomes inaccurate past 5 Gyr. The plot looks qualitatively the same
when varying $M_{tot}$.

\begin{figure}[!ht]
\begin{center}
\includegraphics[scale=0.5]{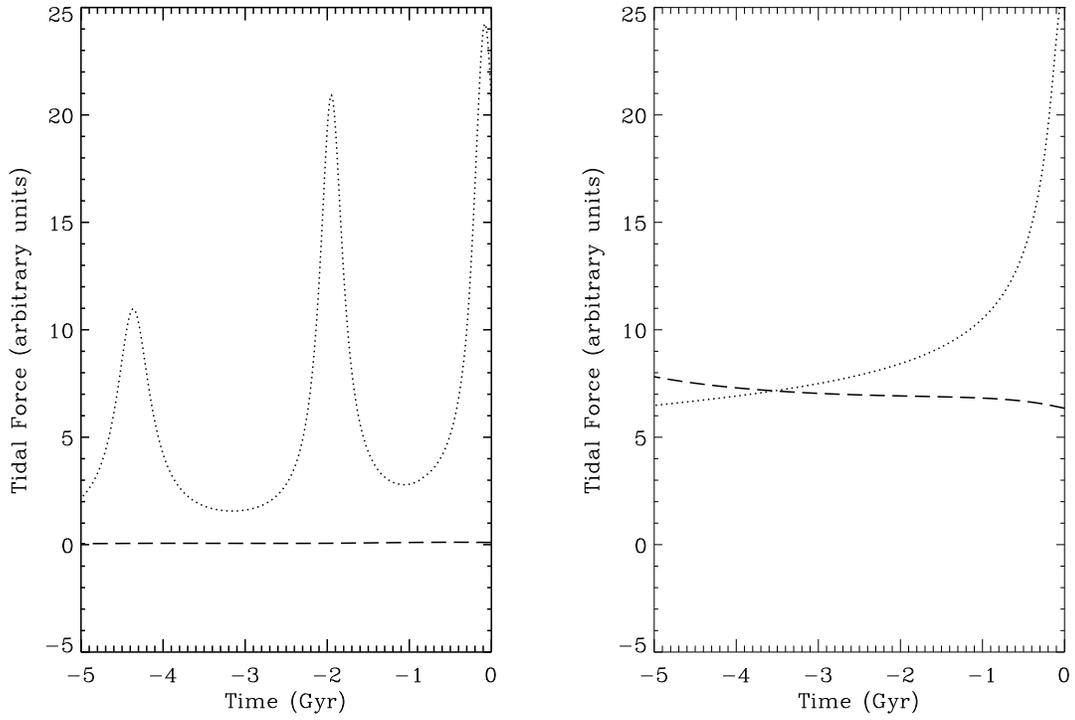}
\caption{\singlespace The tidal force (in arbitrary units) exerted on
the LMC by the MW (dotted line) and M31 (dashed line) as a function of
time. (\textit{Left}) The forces on the LMC orbit that is furthest from
M31, and (\textit{right}) the forces on the LMC orbit that goes closest to
M31.}
\label{tides}
\end{center}
\end{figure}

Figure~\ref{tides} quantifies the strength of the tidal force exerted
on the LMC by the MW and M31 as a function of time. The tidal force is
calculated by twice differentiating the gravitational potential of the
galaxy in question at the position of the LMC, even though strictly
the tidal field is a traceless tensor that requires tangential
derivatives as well. The radial derivative is proportional to the tide
only for a perturbed body of constant size (see Gardiner \& Noguchi
1996) but is instructive here given our methods. Figure~\ref{tides}
(\textit{left}) shows the relative tidal forces on the LMC orbit that
is furthest from M31 (bound to the MW), and on the right for the LMC
orbit that comes closest to M31 (unbound to the MW). The dotted line
shows the tidal force exerted by the MW and the dashed lines shows the
same for M31. The M31 tide takes over for orbits moving $3\sigma$
faster than the mean, but is insignificant for lower values of LMC
velocity.

%Finally, we check to see if this has any effect on the deviation of
%the LMC's past orbit from the location of the MS. In
%Figure~\ref{polarprojM31NFW} we show a polar projection in ($l,b$)
%coordinates for the location of the MS (solid blue line), the past
%orbit of the LMC in the Local Group model (dotted green line) and the
%past orbit in the MW-only NFW model (dashed pink line). There is no
%pronounced change to the orbit, strengthening our assertion in B07
%that it is $\mu_N$ of the proper motion that determines the location
%of the orbit in ($l,b$)-space, regardless of the MW model.

%\begin{figure}[!ht]
%\begin{center}
%\includegraphics[scale=0.7]{polarprojM31NFW.ps}
%\caption{\singlespace Polar projection of the location of the MS
%(solid blue line), the past orbit of the LMC in the Local
%Group$+$intra-group medium model (dotted green line) and the past
%orbit in the MW-only NFW model (dashed pink line).}
%\label{polarprojM31NFW}
%\end{center}
%\end{figure}

\subsection{MW Mass}
If the LMC is indeed on its first passage about the MW then we can use
the timing argument to estimate the total mass of the LMC-MW
system. This is not strictly a correct application of the timing
argument because for a small galaxy such as the LMC, the details of
the neighboring mass distribution become important and cannot be
treated as homogeneous and isotropic. Even so, it is instructive to
see what kind of limits the timing argument places on the mass of the
MW given the assumption that the LMC and MW are encountering each
other for the first time. Since the current position and velocity of
the LMC relative to the MW is known, this specifies the semi-major
axis of the relative orbit via the cubic equation
\begin{equation}
2n^2a^3 - n^2 r_L a^2 - r_L v_L^2 = 0,
\end{equation}
where $a$ is the semi-major axis, $n= 2 \pi /T$ is the mean motion,
and $r_L$ and $v_L$ are the position and velocity at
time zero. If $T = t_{\rm H} = 13.7$ Gyr (Hinshaw \etal 2008), we can
solve uniquely for $\mu = G(M_{\rm MW} + M_{\rm LMC}) = n^2a^3$. A
Monte-Carlo analysis of the errors in LMC velocity gives:
\begin{equation}
M_{\rm MW} + M_{\rm LMC} = (8.7 \pm 0.8) \times 10^{11} M_\odot,
\end{equation} 
which is in good agreement with $\Lambda$CDM-based estimates for the
mass of the MW (e.g., Klypin \etal 2002). It is lower than recent
observational estimates based on VLBA measurements by about a factor
of 2 (Reid \etal 2009).

\section{Conclusions}\label{conclusion}
We have studied the past orbital histories of the Clouds in a model
for the Local Group whose properties are constrained using the timing
argument. This study has been motivated by the substantial increase in
the tangential velocity of the LMC (K1; Piatek \etal 2008;
Kallivayalil \etal 2009) and thus the possibility that the LMC is on
its first passage about the MW (Besla \etal 2007), as well as a new
assessment of the tangential velocity of M31 (VG08). The timing
argument provides a natural framework within which to test the effect
of `maximal' M31 models on the past orbit of the LMC. The goal is to
evaluate whether, as in Raychaudhury \& Lynden-Bell (1989) for
instance, tidal torques exerted by M31 are important for the LMC.

We investigate the effects of $M_{tot}$, $f=M_{M31}/M_{MW}$, M31
proper motion and LMC proper motion. We find that $M_{tot}$ and M31
proper motion both act in the same way : larger values of M31
tangential velocity lead to larger values of $M_{tot}$, period and
pericentric distance of the MW-M31 orbit. These in turn affect the
period and peri/apogalactic distances of the MW-LMC orbit, but within
the accepted error-space, M31 proper motion does not significantly
affect the trajectory of the LMC orbit. $M_{tot}$ does affect the
trajectory of the orbit if M31 is sufficiently massive ($M_{tot} > 3.5
\times 10^{12} M_\odot$) and if the western component of the LMC
proper motion, $\mu_W$ is increased by roughly $3\sigma$. The amount
of this increase depends on $M_{tot}$. However, for the rest of the
explored parameter-space, the influence of M31 is negligible. Furthermore,
it is unlikely that we have underestimated the proper motion of the
LMC by this amount, and thus conclude that the M31 tide is likely
insufficient in generating the high angular momentum of the LMC
orbit. In other words, the angular momentum problem remains.

Finally, motivated by the fact that the LMC could be on its first
passage about the MW, we calculate the implied MW+LMC mass if they
were last co-located a Hubble time ago. A Monte-Carlo analysis of the
errors gives: $M_{\rm MW} + M_{\rm LMC} = (8.7 \pm 0.8) \times 10^{11}
M_\odot$. Since $M_{\rm MW} >> M_{\rm LMC}$, this quantity is roughly
the mass of the MW, and is in good agreement with $\Lambda$CDM
predictions (e.g., Klypin \etal 2002).

It is clear from this analysis that the large tangential motion of the
LMC cannot be easily explained away with manipulations of the Local
Group model. Confirming this large motion is a high priority. In
Kallivayalil \etal (2009) we presented an ongoing analysis of the
proper motions of the Magellanic Clouds using a third epoch of WFPC2
data centered on background quasars. The results so far as consistent
with those presented in K1. At present the RMS error in the position
of the quasar is roughly 3 times as large for WFPC2 as for
ACS. However, with an improved method to deal with charge-transfer
efficiency and magnitude-related effects, and with the increase in
time-baseline from 2 to 5 years, we expect final error bars for the
proper motions that are smaller by a factor of $\sim 2$ from K1.
This, combined with our understanding of the properties of the MS,
will allow us to better constrain the orbit of the Clouds and make
more specific predictions as to their origin.

\acknowledgments 
We would like to thank Ed Bertschinger, Paul Schechter and TJ Cox for useful
discussions.

\newpage
%TABLE 1%%%%%%%%%%%%%%%%%%%%%%%%%%%%%%%
\begin{deluxetable}{llll}
%\tabletypesize{\tiny}
%\rotate
\tablewidth{0pt}
\tablecolumns{4}
\tablecaption{Parameter-space}
\tablehead{
\colhead{Parameter}  & \colhead{Values} & 
\colhead{Step-size} & \colhead{$1\sigma$ error}}
\startdata
$M_{tot}$ ($M_\sun$)................................................. & $2.6-6 \times 10^{12}$ & 
$0.1 \times 10^{12}$ & - \\
$f$................................................................ & $0.8 - 2.0$ & 0.1 & - \\
M31 tangential velocity ($\kms$)............ & 0; 42 & - & $\leq 56$ \\
LMC proper motion ($\mu_N, \mu_W$) $(\masyr)$ & 0.44, -2.03 & Monte Carlo dist. 
& 0.05, 0.08 \\
SMC proper motion ($\mu_N, \mu_W$) $(\masyr)$ & -1.17, -1.16 & - & Fixed \\
\enddata
\end{deluxetable}

\newpage

\newpage
\newpage

\newpage

\newpage

%%%%%%%%%%%%%%%%%%%%%%%%%%%%%%%%%%%%%%%%%%%%%%%%

\end{document}